\documentclass{jetpl}
\twocolumn

\usepackage[english,russian]{babel}

\title{Low-Temperature Mobility of Surface Electrons\\ and Ripplon-Phonon Interaction in Liquid Helium}

\rtitle{Low-Temperature Mobility of Surface Electrons\ldots}

\sodtitle{Low-Temperature Mobility of Surface Electrons and Ripplon-Phonon Interaction in Liquid
Helium}

\author{A.\,I.\,Safonov\/\thanks{safonov@isssph.kiae.ru}, I.\,I.\,Safonova and S.\,S.\,Demukh}

\rauthor{A.\,I.\,Safonov, I.\,I.\,Safonova and S.\,S.\,Demukh}

\sodauthor{Safonov~A.\,I., Safonova~I.\,I. and Demukh~S.\,S.}

\address{Russian Research Centre Kurchatov Institute, pl. Akademika Kurchatova 1, Moscow, 123182 Russia}

\dates{February 4, 2010}{March 4, 2010}

\abstract{The low-temperature dc mobility of the two-dimensional electron system localized above
the surface of superfluid helium is determined by the slowest stage of the longitudinal momentum
transfer to the bulk liquid, namely, by the interaction of surface and volume excitations of
liquid helium, which rapidly decreases with temperature. Thus, the temperature dependence of the
low-frequency mobility is
$\mu_\textrm{dc}\approx8.4\times10^{-11}n_eT^{-20/3}$~cm$^4$K$^{20/3}$/(V$\cdot$s), where $n_e$ is
the surface electron density. The relation $T^{20/3}E_\bot^{-3}\ll2\times10^{-7}$ between the
pressing electric field (in kV/cm) and temperature (in K) and the value $\omega\lesssim10^8\cdot
T^5$~K$^{-5}$s$^{-1}$ of the driving-field frequency have been obtained, at which the above effect
can be observed. In particular, $E_\bot\simeq1$~kV/cm corresponds to $T\lesssim70$~mK and
$\omega/2\pi\lesssim30$~Hz.}


\begin{document}

\maketitle

Low-dimensional systems of charges have been studied theoretically and experimentally for about
four decades. These systems include the two-dimensional electron gas in semiconductor structures
and the electron systems localized above the surface of liquid helium \cite{SM_1989, Edelman}. The
relaxation times of the excited spin or Rydberg states of surface electrons turn out to be much
longer owing to a weak coupling to the helium surface. This makes the surface electrons (SE) quite
attractive as possible memory units of a quantum computer~\cite{Dykman}. At the same time, the
relative weakness of the relaxation mechanisms determines the ultimately high mobility
$\mu\sim10^4$~m/V$\cdot$s of the SEs, probably the highest of all known two-dimensional systems.

At $T<0.7$~K, when the saturation pressure of helium vapor is exponentially small, the SE mobility
is limited by their interaction with ripplons, the elementary excitations of the helium surface,
whereas the direct electron-phonon interaction is ineffective. All previous theoretical and
experimental works on the SE mobility assumed that the longitudinal momentum of the electrons
transferred due to their interaction with ripplons is immediately accumulated by the entire liquid
(see, e.g., \cite{SM_1989}. However, there are indications that in fact the relaxation of the
longitudinal momentum of the electrons proceeds in two stages, via the momentum transfer to
ripplons and further from the ripplons to the phonons of the bulk liquid, and is limited by the
slowest stage.

Generally speaking the experimental observation of the ripplon-phonon interaction is a separate
and quite formidable problem. One of possible solutions is to measure the heat transfer along the
helium surface \cite{MEN}. However, in this case, one has to use thin films to avoid a
giant-effective thermal conductance of a bulk liquid, which could easily mask the surface heat
transfer. On the other hand, in the case of films, an important role is played by the surface
smoothness, as the ripplons can be scattered not merely by phonons but also by surface
imperfections caused by substrate defects (roughness). Such a one-body elastic scattering of
ripplons has a much weaker temperature dependence than the ripplon-phonon interaction and
therefore dominates at low temperature \cite{ripplon}. Guaranteed exclusion of this one-body
elastic channel of the ripplon flow dissipation imposes quite severe requirements on the surface
smoothness over a large area. Thus, the contribution of the ripplon-phonon processes to the
thermal conductance of helium surface turns out to be barely detectable. Roche et al.
\cite{Roche95} implemented capacitive excitation and detection to measure the damping of
long-wavelength ripplons. However, this technique is hardly applicably for studying thermal
ripplons whose wavelength is just ~10 nm, i.e., three orders of magnitude shorter than in
experiments of Roche et al. In this work, we show that the use of surface electrons might be very
promising.

The surface (ripplons) and volume (phonons) quasiparticles of liquid helium interact only to the
extent of a small compressibility of helium. The rate of the dominant process of momentum transfer
at low temperatures (the phonon creation with the annihilation of two ripplons \cite{RSS}) is
proportional to $T^{20/3}$~\cite{ripplon}. Consequently, the ripplon subsystem in the
$T\rightarrow0$ limit appears to be practically isolated from the liquid bulk and has to follow
the regular motion of the electrons. Thus, the SE mobility should infinitely increase with a
decrease in temperature. However, nothing of that kind was observed in experiments at least down
to 20 mK~\cite{Shirahama, Kono}. On the contrary, the mobility of the electrons, which form a
Wigner solid \cite{Wigner, Grims_WS} at low temperature, even somewhat decreases and levels off at
$T\rightarrow0$. This work is aimed at resolving this contradiction, finding the conditions, under
which the SE mobility is determined by the ripplon-phonon interaction (or, at least, when the
respective contribution is considerable), and calculating the dependence of the mobility on
temperature, the frequency $\omega$ of the driving electric field
$E_\parallel(t)=E_\parallel\textrm{e}^{i\omega t}$ and the magnitude of the pressing field
$E_\perp$.

We suggest that the said contradiction is caused by the fact that the measurements are usually
conducted in the ac regime at a relatively high frequency and that the electron effective mass is
several orders of magnitude less than that of ripplons. As a result, the amplitude of the in-plane
oscillations of the ripplons subsystem is vanishingly small. A mechanical analogue of this
situation could be a small rubber on the surface of a massive plate, which in turn lies on a
smooth ice. The in-plane motion of the rubber driven by a periodic force is detected. It is
intuitively clear that in order to observe the joint motion of electrons and ripplons relative to
the bulk liquid one has to perform the measurements in the dc regime or at a very small frequency.
It is also clear that one should aim at increasing the electron-to-ripplon ratio of the effective
mass densities.

We shall consider the ripplon and electron subsystems separately. This requires that the
equilibration time within each subsystem be substantially shorter than the relaxation time between
the subsystems. Next, we do not include the effect of the transverse magnetic field, which is
typically used in the mobility measurement in the Corbino geometry but absolutely insignificant
for the present problem. The amplitude $E_\parallel$ of the driving electric field is thought to
be sufficiently small so that the effects associated with the electron overheating could be
neglected. We also disregard the boundary effects assuming the infiniteness of the two-dimensional
system. Taking into account the above remarks, the equations of motion of electrons and ripplons
in the $\tau$ approximation read as
\begin{equation}\label{Eq1}
\frac{d{\bf u}_e}{dt} = -\frac{e{\bf E}_\parallel}{m^\ast}+\frac{{\bf u}_{\rm R}-{\bf
u}_e}{\tau_e}
\end{equation}
and
\begin{equation}\label{Eq2}
\frac{d{\bf u}_{\rm R}}{dt} = -\gamma\frac{{\bf u}_{\rm R}-{\bf u}_e}{\tau_e}-\frac{{\bf u}_{\rm
R}}{\tau_{\rm RP}},
\end{equation}
respectively. Here, ${\bf u}_e$ (${\bf u}_{\rm R}$) is the in-plane velocity of the electron
(ripplon) subsystem as a whole, $e$ is the unit charge, $m^\ast$ and $n_e$ are the the effective
mass and number density of electrons, $\rho_{\rm R}=1.67\times10^{-10}T^{5/3}~$K$^{-5/3}$g/cm$^2$
is the mass density associated with ripplons, $\gamma=n_em^\ast/\rho_{\rm R}$ is the
electron-to-ripplon mass-density ratio. The electron momentum-relaxation time $\tau_e$ is related
to the SE mobility $\mu$ with respect to ripplons as $\tau_e=m^\ast\mu/e$ (see below) and weakly
depends on temperature in the range under consideration. According to the experimental data on the
electron mobility \cite{Dotsenko} and effective mass \cite{Syvokon}, the electron-ripplon
relaxation time at $T=70\div80$~mK and $n_s=1.3\times10^9$~cm$^{-2}$ is
$\tau_e\sim5\times10^{-7}$~s. Thus, at $T\lesssim0.1$~K the ripplon-phonon momentum relaxation
time $\tau_{\rm RP}=8.8\times10^{-9}T^{-5}$~K$^5$s~\cite{ripplon} is much greater than $\tau_e$.
Note, that the comprehensive kinetic calculation of $\tau_e$ for free electrons is also based on
the equilibrium of the electron and ripplon subsystem \cite{SM_1989}. In particular, according to
\cite{SM_1989}, the rate of electron-electron collisions in a typical experiment is much higher
than $\tau_e^{-1}$.

In a weak pressing field $E_\perp$, the effective mass of electrons almost coincides with their
bare mass, $m^\ast\approx m$. In this case, the electrons appear to be orders of magnitude
``lighter'' than the ripplons ($\gamma\ll1$) and the latter do not feel the electron motion,
$u_{\rm R}\ll u_e$. In this case, Eq. (\ref{Eq1}) yields the usual expression for the electron
mobility (to be compared with Eq. (10.7) of \cite{SM_1989})
\begin{equation}\label{Eq3}
\mu\equiv-\frac{{\rm
Re}~u_e}{E_\parallel}=\left(\frac{e}{m^\ast}\right)\frac{\tau_e}{1+\omega^2\tau_e^2}.
\end{equation}
At not to high driving-field frequency $\omega\tau_e\ll1$ (which is equivalent to neglecting the
left-hand side of Eq. (\ref{Eq1})), we find $\mu=e\tau_e/m^\ast$ \cite{tau}.

In a strong pressing field the picture is substantially changed. First of all, the maximum
achievable electron density increases proportional to the pressing field, $E_\perp=2\pi en_e^{\rm
max}$. In addition, the electron effective mass as such increases by two or even three orders of
magnitude. In particular, $m^\ast/m$ increases by from 500 to 5000 nearly proportional to the
square of the pressing field $E_\perp=0.57\div1.672$~kV/cm at a constant density
$n_e=6.3\times10^8$~cm$^{-2}$ \cite{Stan} and slightly decreases (from ~2000 to 1500) with an
increase in density from 5 to $12\times10^8$~cm$^{-2}$ at a constant pressing field
$E_\perp=1.15$~kV/cm~\cite{Syvokon}. An increase in the effective mass is caused by the fact that
the strong field presses electrons into the helium surface, thus forming dimples. Therefore, the
electron motion along the surface involves a considerable mass of helium associated with the
displacement of the dimples. Though the resultant mass densities of electrons and ripplons are
still substantially different, this mass increase may be sufficient to fulfill the condition
$\tau_e\ll\gamma\tau_{\rm RP}$, which is equivalent to $\mu/en_e\ll\tau_{\rm RP}/\rho_{\rm
R}\simeq53~T^{-20/3}~$K$^{20/3}$s$\cdot$cm$^2$/g. For example, at $T=70$~mK, $n_e=10^9$~cm$^{-2}$
and $E_\perp=1$~kV/cm one has $\tau_e\simeq5\times10^{-7}$~s, $\gamma\simeq3\times10^{-3}$,
$\tau_{\rm RP}\simeq5\times10^{-3}$~s and, consequently, $\gamma\tau_{\rm
RP}\sim1.5\times10^{-5}$~s. In this case, we find from Eq. (\ref{Eq2}) with $\omega\tau_e\ll1$
that the electron and ripplon velocities almost coincide, $u_{\rm R}\approx u_e\equiv u$, and,
consequently,
\begin{equation}\label{Eq4}
u\approx-eE_\parallel\frac{n_e\tau_{\rm RP}}{\rho_{\rm R}}\left(\frac{1+i\omega\tau_{\rm
RP}}{1+\omega^2\tau_{\rm RP}^2}\right),
\end{equation}
where we neglected $\gamma$ with respect to unity.

As expected, the amplitude $|u|$ of the velocity oscillations has a maximum at zero frequency of
the driving field and decreases as $1/\omega$ at high frequencies ($\omega\tau_{\rm RP}\gg 1$)
with a $\pi/2$ phase lag compared to the driving field. The appearance of the out-of-phase
component of the electron velocity (and, consequently, the current) serves as the direct evidence
of the discussed effect. This component has a maximum at $\omega\tau_{\rm RP}=1$, which is exactly
one half of the maximum value of the velocity. Equation (\ref{Eq4}) with $\omega=0$ yields the dc
electron mobility relative to the bulk liquid
\begin{equation}\label{Eq5}
\mu_\textrm{dc} = \frac{en_e\tau_{\rm RP}}{\rho_{\rm
R}}\approx8.4\times10^{-11}n_eT^{-20/3}\frac{\textrm
{cm}^4\textrm{K}^{20/3}}{\textrm{V}\cdot\textrm{s}}.
\end{equation}
The found temperature dependence of the low-frequency mobility for $n_e=1.4\times10^7$,
$1.08\times10^8$ and $1.3\times10^9$~cm$^{-2}$ ($E_\perp=1.25$ kV/cm) is shown in figure by the
dotted, dashed and solid line, respectively. The results of mobility measurements via the
longitudinal ac conductivity $\sigma_{xx}$ of the two-dimensional system in the Corbino geometry
at $\omega/2\pi=10$~kHz, $E_\perp=15$~V/cm, $n_e=1.4\times10^7$~cm$^{-2}$ \cite{Kono};
$\omega/2\pi=10$~kHz, $E_\perp=92.5$~V/cm, $n_e=1.08\times10^8$~cm$^{-2}$ \cite{Shirahama}, and
via the damping of the mixed phonon-ripplon modes of a Wigner solid at $\omega/2\pi=5\div10$~MHz,
$n_e=1.3\times10^9$~cm$^{-2}$ \cite{Dotsenko} are also shown for comparison.

\begin{figure}
\includegraphics[width=8.5cm]{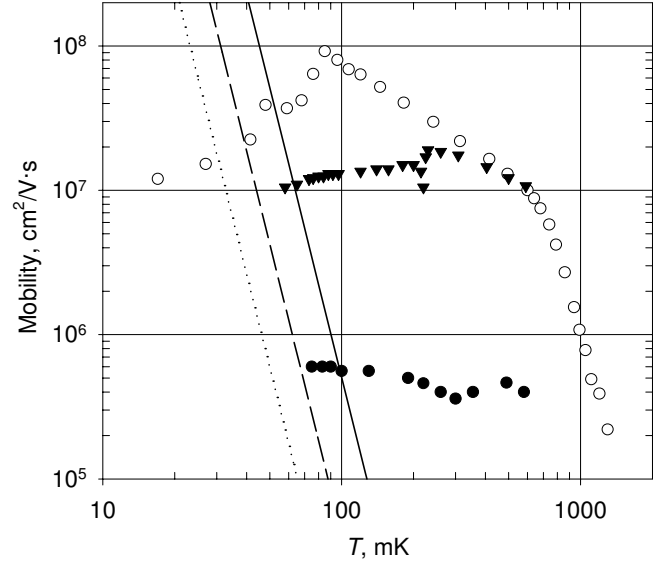}
\caption{Temperature dependence of the SE mobility measured via the longitudinal ac conductivity
$\sigma_{xx}$ of the two-dimensional system in the Corbino geometry at ($\circ$)
$\omega/2\pi=10$~kHz, $E_\perp=15$~V/cm, $n_e=1.4\times10^7$~cm$^{-2}$ \cite{Kono};
($\blacktriangledown$) $\omega/2\pi=10$~kHz, $E_\perp=92.5$~V/cm, $n_e=1.08\times10^8$~cm$^{-2}$
\cite{Shirahama}, and ($\bullet$) via the damping of the mixed phonon-ripplon modes of a Wigner
solid at $\omega/2\pi=5\div10$~MHz, $n_e=1.3\times10^9$~cm$^{-2}$ \cite{Dotsenko}. The straight
lines show the dc mobility (\ref{Eq5}) for the respective values of $n_e$ (from left to right in
the order of increasing density). Some experimental points are omitted for better visibility.}
\label{Fig}
\end{figure}

The theoretical dc mobility of free electrons relative to ripplons in a strong pressing field
is~\cite{Shikin}
\begin{equation}\label{Eq6}
\mu_0=\frac{8\alpha\hbar}{e^2E_\perp^2}\frac{e}{m}=\frac{9.3\times10^{11}\textrm{V/s}}{E_\perp^2},
\end{equation}
where $\alpha=0.378$ erg/cm$^2$ is the surface tension of helium. Thus, in full agreement with
experiments~\cite{Grims1976, Platzman_Beni, Eselson}, the mobility is inversely proportional to
the square of the pressing field and does not depend on temperature. In the case of a Wigner
solid, the behavior of the mobility changes substantially, first of all, because, depending on the
magnitude $E_\parallel$ and $E_\perp$ of the pressing and driving field, excitation frequency and
temperature, the two-dimensional electron system may find itself in a so-called coupled
phonon-ripplon (CPR) state formed by the electrons and the respective dimples on the helium
surface or slide over the helium surface \cite{Shirahama_PRL, Kono_SS}. Under the transition
between these two states, the electron mobility exhibits an abrupt change, which may reach an
order of magnitude. In the CPR state of interest, which in fact appears in strong pressing fields
and a weak driving field and at low frequencies, the dependence $\mu\propto E_\perp^{-2}$ is
generally preserved \cite{Stan, Mehrotra} and the mobility of a Wigner solid also weakly depends
on temperature \cite{Kono, Dotsenko, Stan}. Taking for the estimate the value
$\mu=6\times10^5$~cm$^2$/V$\cdot$s obtained at $n_e=1.3\times10^9$~cm$^{-2}$ and $T=70$~mK
\cite{Dotsenko}, we find that the rates of momentum transfer from electrons to ripplons and from
ripplons to photons of helium become equal at $T\simeq98$~mK (see figure). For the ripplon-phonon
interaction to be the bottleneck of the longitudinal momentum relaxation the measurements should
be carried out at $T\lesssim70$~mK. According to Eq. (\ref{Eq4}), the frequency $\omega$ of the
driving field in this case should not be greater than
$\tau_\textrm{RP}^{-1}\simeq1.1\times10^8T^5$~K$^{-5}$s$^{-1}$, which corresponds to
$\omega/2\pi\lesssim30$~Hz at the specified temperature.

Thus, we have seen that observing the ripplon-phonon interaction in liquid helium via the SE
mobility requires the simultaneous fulfillment of the conditions
\begin{equation}\label{neq1}
\tau_e\ll\frac{m^\ast n_e}{\rho_{\rm R}}\tau_{\rm RP}
\end{equation}
and
\begin{equation}\label{neq2}
\omega\tau_{\rm RP}\lesssim1.
\end{equation}
Taking into account the dependence of the effective mass $m^\ast$ and limiting density $n_e$ of
the surface electrons on the pressing field $E_\bot$ and the temperature dependence of $\rho_{\rm
R}$ and $\tau_{\rm RP}$, these conditions may be rewritten in a more practical form
\begin{equation}\label{neq11}
T^{20/3}E_\bot^{-3}\ll2\times10^{-7},
\end{equation}
\begin{equation}\label{neq21}
\omega/2\pi\lesssim1.5\times10^7T^5~\textrm{Hz},
\end{equation}
where the temperature, field and frequency are expressed in Kelvin, kV/cm and Hz, respectively.

In experiments of Dotsenko et al. \cite{Dotsenko}, the condition (\ref{neq11}) was fulfilled.
However, the condition (\ref{neq21}) was violated so that the relative contribution of the
ripplon-phonon interaction in helium to the mobility of the electron crystal was no more than
$\sim$10$^{-5}$. The technique of mobility measurement via the damping of hybrid phonon-ripplon
modes of a Wigner solid seems to be fundamentally inapplicable for observing the effect under
consideration, since the minimum frequency of the phonon-ripplon modes is given by the lattice
constant of the electron crystal and the dispersion relation of ripplons and therefore cannot be
smaller than $\sim(\alpha/\rho)^{1/2}n_e^{3/4}$, where $\rho=0.145$ g/cm$^3$ is the mass density
of $^4$He at $T=0$ \cite{Deville}. The mobility of the two-dimensional electron system can be also
determined from its longitudinal conductivity $\sigma_{xx}$ measured in the Corbino geometry.
However, the respective experiments \cite{Shirahama, Kono} violated both conditions, primarily
because of the smallness of the pressing field. In particular, at $E_\bot=92.5$~V/cm the
inequalities (\ref{neq11}), (\ref{neq21}) yield $T\leq25$~mK and $\omega/2\pi\lesssim0.14$~Hz
whereas the measurements were carried out at 10 and 100 kHz.

Thus, we have shown that the ripplon-phonon interaction in superfluid helium can be studied with
the use of surface electrons and determined the conditions to be satisfied by the experimental
parameters, i.e., temperature, excitation frequency, the density of the two-dimensional electron
system, and the magnitude of the pressing electric field.

We are truly grateful to S.S.Sokolov and V.E.Sivokon' for fruitful discussions, to K.Kono for
providing us with some of his relevant publications and to I.S.Yasnikov for reading the manuscript
and valuable remarks. This work was supported by the Russian Foundation for Basic Research,
project no. 09-02-01002-а.

\vfill\eject


\begin{thebibliography}{9}
\bibitem{SM_1989}V. B. Shikin and Yu. P Monarkha, {\it Two-Dimensional Charged Systems in Helium} (Moscow, Nauka,
1989) (in russian).
\bibitem{Edelman}V. S. Edelman, Sov. Phys. Uspekhi {\bf23}, 227 (1980) [Usp. Fiz. Nauk. {\bf130}, 675 (1980)].
\bibitem{Dykman}M. I. Dykman, P. M. Platzman and P. Seddighrad, Phys. Rev. B {\bf67}, 155402
(2003).
\bibitem{MEN}I.\,B. Mantz, D.\,O. Edwards and V.\,U. Nayak, Phys. Rev. Lett. {\bf 44}, 66 (1980);
Errata {\bf 44}, 1094 (1980).
\bibitem{ripplon}A.\,I. Safonov, S.\,S. Demukh and A.\,A. Kharitonov, JETP Lett. {\bf79}, 304 (2004) [Pis'ma Zh. Eksp. Teor. Fiz. {\bf79}, 362 (2004)].
\bibitem{Roche95}P. Roche, G. Deville, K.\,O. Keshishev, N.\,J. Appleyard, and F.\,I.\,B.
Williams, Phys. Rev. Lett. {\bf 75}, 3316 (1995).
\bibitem{RSS}M.\,W. Reynolds, I.\,D. Setija and G. V. Shlyapnikov, Phys. Rev. {\bf B 46}, 575 (1992).
\bibitem{Shirahama}K. Shirahama, S. Ito, H. Suto, and K. Kono, J. Low Temp. Phys. {\bf101}, 439 (1995).
\bibitem{Kono}K. Kono and K. Shirahama, in {\it Two-Dimensional Electron Systems}, ed. E.Y.Andrei
(Kluwer Acad. Publ., 1997), p.p. 175-189.
\bibitem{Wigner}E. Wigner, Trans. Faraday Soc. {\bf34}, 678 (1938).
\bibitem{Grims_WS}C. C. Grims and G. Adams, Phys. Rev. Lett. {\bf42}, 795 (1979).
\bibitem{Dotsenko}V. V. Dotsenko, V. E. Sivokon', Yu. Z. Kovdrya, and V. N. GrigorТev, Low Temp. Phys. {\bf23}, 772 (1997) [Fiz. Nizk. Temp. {\bf23}, 1028 (1997)].
\bibitem{Syvokon}V. E. Syvokon, Yu. Z. Kovdrya, and K. A. Nasyedkin, J. Low Temp. Phys. {\bf144}, 35 (2006).
\bibitem{tau}We disregard a factor-of-two difference in $\tau_e$ at high- and low-frequency excitation
\cite{Platzman_Beni} because of the approximate character of the inequality (\ref{neq1}), which
includes $\tau_e$.
\bibitem{Stan}M. A. Stan and A. J. Dahm, Phys. Rev. B {\bf40}, 8995 (1989).
\bibitem{Shikin}V. B. Shikin and Yu. P. Monarkha, J. Low Temp. Phys. {\bf16}, 193 (1974).
\bibitem{Grims1976}C. C. Grims and G. Adams, Phys. Rev. Lett. {\bf36}, 145 (1976).
\bibitem{Platzman_Beni}P. M. Platzman and G. Beni, Phys. Rev. Lett. {\bf36}, 626 (1976).
\bibitem{Eselson}B. N. Esel'son, A. S. Rybalko and S. S. Sokolov, Sov. J. Low Temp. Phys. {\bf6}, 544 (1980) [Fiz. Nizk. Temp. {\bf6}, 1120 (1980)].
\bibitem{Shirahama_PRL}K. Shirahama and K. Kono, Phys. Rev. Lett. {\bf74}, 781 (1995).
\bibitem{Kono_SS}K. Kono and K. Shirahama, Surf. Sci. {\bf 361/362}, 826 (1996).
\bibitem{Mehrotra}R. Mehrotra, C. J. Guo, Y. Z. Ruan, D. B. Mast, and A. J. Dahm, Phys. Rev. B {\bf29},
5239 (1984).
\bibitem{Deville}G. Deville, J. Low. Temp. Phys. {\bf72}, 135 (1988).
\end{thebibliography}
\end{document}